\documentstyle[epsfig]{basi}
\begin{document}
\title[NIR Stellar Spectral Library]{A Near-Infrared Stellar Spectral Library: III. J-Band Spectra}
\author[Arvind C. Ranade et al.]%
       {Arvind C. Ranade$^1$, N M Ashok $^2$, Harinder P. Singh$^3$ and
       Ranjan Gupta$^4$\thanks{e-mail:rag@iucaa.ernet.in} \\
 $^1$ Vigyan Prasar, A-50, Institutional Area, Sector-62, NOIDA 201 307, India\\
 $^2$ Physical Research Laboratory, Navrangpura, Ahmedabad 380 009, India \\
 $^3$ Department of Physics \& Astrophysics, University of Delhi, Delhi 110 007, India\\
 $^4$ IUCAA, Post Bag 4,Ganeshkhind, Pune 411 007, India}

%\pubyear{2001}
%\volume{29}
%\setcounter{page}{17}
%\date{Received 2001 May 30; accepted 2001 June 07}
\maketitle
\label{firstpage}
\begin{abstract}This paper is the third in the series of papers published
on near-infrared (NIR) stellar spectral library by Ranade et al.
(2004 \& 2007). The observations were carried out with 1.2 meter
Gurushikhar Infrared Telescope (GIRT), at Mt. Abu, India using a
NICMOS3 HgCdTe $256 \times 256$ NIR array based spectrometer. In
paper I (Ranade et al. 2004), H-band spectra of 135 stars at a
resolution of $\sim 16$\AA~ \& paper II (Ranade et al. 2007), K band
spectra of 114 stars at a resolution of $\sim 22$\AA~ were
presented. The J-band library being released now consists of 126
stars covering spectral types O5--M8 and luminosity classes I--V.
The spectra have a moderate resolution of $\sim 12.5$\AA~ in the J
band and have been continuum shape corrected to their respective
effective temperatures. The complete set of library in near-infrared
(NIR) will serve as a good database for researchers working in the
field of stellar population synthesis. The complete library in J, H
\& K is available online at:
http://vo.iucaa.ernet.in/$\sim$voi/NIR\_Header.html

\end{abstract}

\begin{keywords}
infrared: stars --- techniques: spectroscopic
\end{keywords}
\section{Introduction}
The development in size and quantum efficiency of detectors has completely
revolutionized the field of near infrared astronomy. Due to this several
wide-field surveys like Two Micron All Sky Survey (2MASS; Skrutskie et al. 1997) and
Deep Near Infrared Southern Sky Survey (DENIS; Epchtein et al. 1997) were possible.
Near infrared spectra are useful in many astrophysical applications including
spectral classification, spectral definition of sub dwarf objects, calibration
of temperatures of late-type stars, definition of the end of main sequence
etc. which are currently not well understood.

    The characterization and analysis of stellar infrared spectra is an essential
tool in understanding the physical and chemical processes taking
place in stellar atmosphere (Heras et al. 2002). At the same time
one needs to have accurate modeling of the NIR spectral range, which
in turn must rely on NIR libraries of all types of stars. The first
library of such stellar spectra was published in 1970 by Johnson \&
Me\'ndez; for 32 stars in 1 to 4 $\mu$ m region with the resolving
power varying from 300 to 1000.  A number of atlases at medium
resolution in H and K band are provided by Kleinman \& Hall (1986),
Lan\c{c}on \& Rocca-Volmerange (1992), Origlia et al. (1993), Ali et
al. (1995), Dallier et al. (1996), Ramirez et al. (1997). The more
recent libraries including work of Meyer \& Wallace et al. (1997 \&
1998) for H and K libraries are summarized by Ivanov et al. (2004).
The spectral library of late type stars by Ivanov et al. (2004) has
218 red stars spanning a range of [Fe/H] $\sim$ -2.2 to $\sim$ +0.3
but is not flux calibrated.

   In contrast, J-band is the least explored region of near infrared spectroscopy.
Since the hot stars up to A4 do not have many features in the J band
region (Malkan et al. 2002) the atlases in J band region cover the
cooler part of HR diagram. Some examples of the existing J band
spectral atlases with resolution varying from 1000 to 2500 are that
of evolved stars of S, C \& M types by Joyce et al. (1998), L \& T
dwarfs by McLean et al. (2000), M, L \& T dwarfs by Cushing et al.
(2005) and M2.5 to T6 dwarfs by McLean et al. (2003).

    A library covering the samples over the HR diagram could be the reasonable way
to get the relation of temperature and stellar features. There are
very few libraries in J band which have the complete coverage of HR
diagram in temperature, gravity and metallicity. The library of
Wallace et al. (2000) with 88 sample covering O7 to M6 and I to V
luminosity class with R $\sim$ 3000 and Malkan et al. (2002) with
105 stars from O9.5 to M7 and I-V luminosity which has R $\sim$ 400.
Though stellar spectral classification is easiest to do with high
resolution data, lower resolution is necessary for observations of
substantial number of objects (Malkan et al. 2002). They have
demonstrated that the low resolution data can be used for stellar
classification, since several features depend on the effective
temperature and gravity. In this paper, we present a spectral
library of 126 star in J-band at moderate resolution of 12.5 \AA~
covering larger range in T$_{eff}$ and larger database as compared
to Wallace et al. (2000) and Malkan et al. (2002). In this paper,
$\S 2$ describes the observations and related issues. In $\S 3$, we
describe the basis of selection of the stars for this library and in
$\S 4$ we describe the data reduction and calibration procedure.
Lastly, in $\S 5$ we show examples of some J band spectra and their
comparison with the existing database of Wallace et al. (2000).

\section{OBSERVATIONS}

The database of 126 stars selected in this library were observed in
six different runs from January-April 2003. The details of the log
is shown in Table 1 in which first column gives observing dates and
month, column 2 gives the total number of programme stars observed in each run
and last column gives the total number of standard
stars observed in each run.
\begin{table}
\caption{Observations Log at GIRT}
\begin{tabular}{ccc} \hline
%\begin{tabular}{ccccc}
Dates of Observations &{Program Stars}  &
 Standard Stars\\ \hline

20-24 Jan 03 & 18 & 1 \\
07-12 Feb 03 & 40 & 3 \\
02-04 Mar 03 & 13 & 2 \\
17-20 Mar 03 & 28 & 1 \\
04-07 Apr 03 & 26 & 9 \\
27-30 Apr 03 & 20 & 18 \\ \hline

\end{tabular}
\end{table}
All the observations have been done from the 1.2 meter Gurushikhar
Infrared Telescope (GIRT) of Mt.Abu Infrared Observatory, India
(24$^{0}$39$^{\prime} $ 10.9$^{\prime\prime}$N,
72$^{0}$46$^{\prime}$45.9$^{\prime\prime}$E at an altitude of 1680
meters). The J band long slit spectra were taken from the NIR
Imager/Spectrometer equipped with a 256$\times$256  HgCdTe NICMOS3
array. The slit width corresponds to 2 arc-seconds for the f/13
Cassegrain focus with the slit covering most of 240 arc-seconds
field of view and oriented along North-South direction in the sky.
The reflection grating has 149 lines per mm and is blazed for H band
center wavelength of 1.65 $\mu m$ in the first order and combined
with the slit width of 76 $\mu m$ gives a moderate resolution of
1000. The exposure time for individual spectrum ranged from 1 sec to
120 sec depending on the J magnitude of the program star resulting
in S/N ratio of 50 or better. Two sets of spectra were obtained at
two dithered positions on the array, the typical separation was
about 20 arc-sec. The details of procedure to acquire the data from
the Mt. Abu observatory is discussed in paper I.

 For a majority of the program stars, we have observed a nearby
main-sequence A type star at nearly same air-mass to minimize the
effects of atmospheric extinction. To optimize the observing
efficiency, a single standard star has been observed whenever some
of the program stars happened to be in the nearby region of the sky.
For the early February and late April 2003 observing runs, late B
type standards have been observed. The list of standard stars that
have been observed are given in Table 2. In this table the standard
star identifier with HD number is given in column (1), HR number in column
 (2) and right ascension and declination  for J2000.0 in column (3) and (4) respectively.
Columns (5), (6) and (7) contain the spectro-luminosity class, observed V
magnitude and T$_{eff}$ respectively.
      The wavelength calibration has been performed using telluric
absorption features.
\begin{table}
\caption{Standard Star list with Observational Parameters$^*$}
\begin{tabular}{lccccccl} \hline
HD & HR & $\alpha$(J2000.0)&
 $\delta$(J2000.0) & Sp. Type &
 V$_{mag}$ &
 {T$_{eff}$} ($^{\circ}$K) \\
(1) & (2) & (3) & (4) &
(5) & (6) & (7) \\ \hline

HD028319 & HR1412 & 04 28 39.74 & +15 52 15.17 & A7III & 3.41 & 8150  \\
HD047105 & HR2421 & 06 37 42.70 & +16 23 57.31 & A0IV  & 1.90 & 9520  \\
HD060179 & HR2891 & 07 34 35.90 & +31 53 18.00 & A1V   & 1.58 & 9230  \\
HD065456 & HR3113 & 07 57 40.11 & -30 20 04.46 & A2Vvar& 4.79 & 8970  \\
HD071155 & HR3314 & 08 25 39.63 & -03 54 23.13 & A0V   & 3.90 & 9520  \\
HD079469 & HR3665 & 09 14 21.86 & +02 18 51.41 & B9.5V & 3.88 & 10010 \\
HD082621 & HR3799 & 09 34 49.43 & +52 03 05.32 & A2V   & 4.48 & 8970  \\
HD085235 & HR3894 & 09 52 06.36 & +54 03 51.56 & A3IV  & 4.56 & 8720  \\
HD087737 & HR3975 & 10 07 19.95 & +16 45 45.59 & A0Ib  & 3.51 & 9730  \\
HD087901 & HR3982 & 10 08 22.31 & +11 58 01.95 & B7V   & 1.35 & 13000 \\
HD094601 & HR4259 & 10 55 36.82 & +24 44 59.30 & A1V   & 4.50 & 9230  \\
HD097633 & HR4359 & 11 14 14.41 & +15 25 46.45 & A2V   & 3.32 & 8970  \\
HD103287 & HR4554 & 11 53 49.85 & +53 41 41.14 & A0Ve  & 2.43 & 9520  \\
HD106591 & HR4660 & 12 15 25.56 & +57 01 57.42 & A3V   & 3.30 & 8720  \\
HD118098 & HR5107 & 13 34 41.60 & -00 35 44.95 & A3V   & 3.40 & 8720  \\
HD130109 & HR5511 & 14 46 14.92 & +01 53 34.39 & A0V   & 3.72 & 9520  \\
HD139006 & HR5793 & 15 34 41.27 & +26 42 52.90 & A0V   & 2.21 & 9520  \\
HD141003 & HR5867 & 15 46 11.26 & +15 25 18.57 & A2IV  & 3.66 & 8970  \\
HD153808 & HR6324 & 17 00 17.37 & +30 55 35.06 & A0V   & 3.91 & 9520  \\
HD155125 & HR6378 & 17 10 22.69 & -15 43 29.68 & A2.5Va& 2.43 & 8845  \\ \hline
\end{tabular}
\\
$^*$ (3)-(6) From SIMBAD database,(7) From Lang (1992)
\end{table}

\section{SELECTION OF STARS}

While building a spectral library, it is very important  that one
includes various spectral types so that we have a homogeneous and
comprehensive coverage of all possible spectro-luminosity classes.
To optimize the observing efficiency stars up to a magnitude of V
$\sim$ 7 were selected for the present programme. The histogram in
Fig. 1 represents the total number of stars covered in terms of
spectral types (top panel) and luminosity classes (bottom panel).
The details of number of stars covered in terms of spectral types
per luminosity class is illustrated by the histogram in Fig. 2. It
may be noted that we have covered the HR diagram in effective
temperature and luminosity parameters reasonably well, although we
do not have enough stars for luminosity class II and main sequence
spectral type O. The details of program stars along with the NIR
magnitudes in J, H, K, L \& M bands are listed in Table 3. In this
table, the first column contains the program star ID, columns (2) to
(6) list the J, H, K, L \& M magnitudes respectively. The references
from which they have been taken are listed in column 7.

The detailed criteria for the selection of stars with their
references are discussed in paper I. We have covered a reasonable
region of parameter space in temperature, gravity and metallicity.
Fig. 3 shows the plot of {\it log g} vs. T$_{eff}$ for the GIRT
stars. Figs. 4 \& 5 shows the plot of [Fe/H] vs. T$_{eff}$ and {\it
log g} respectively for the GIRT stars.

\begin{figure}
\epsfig{file=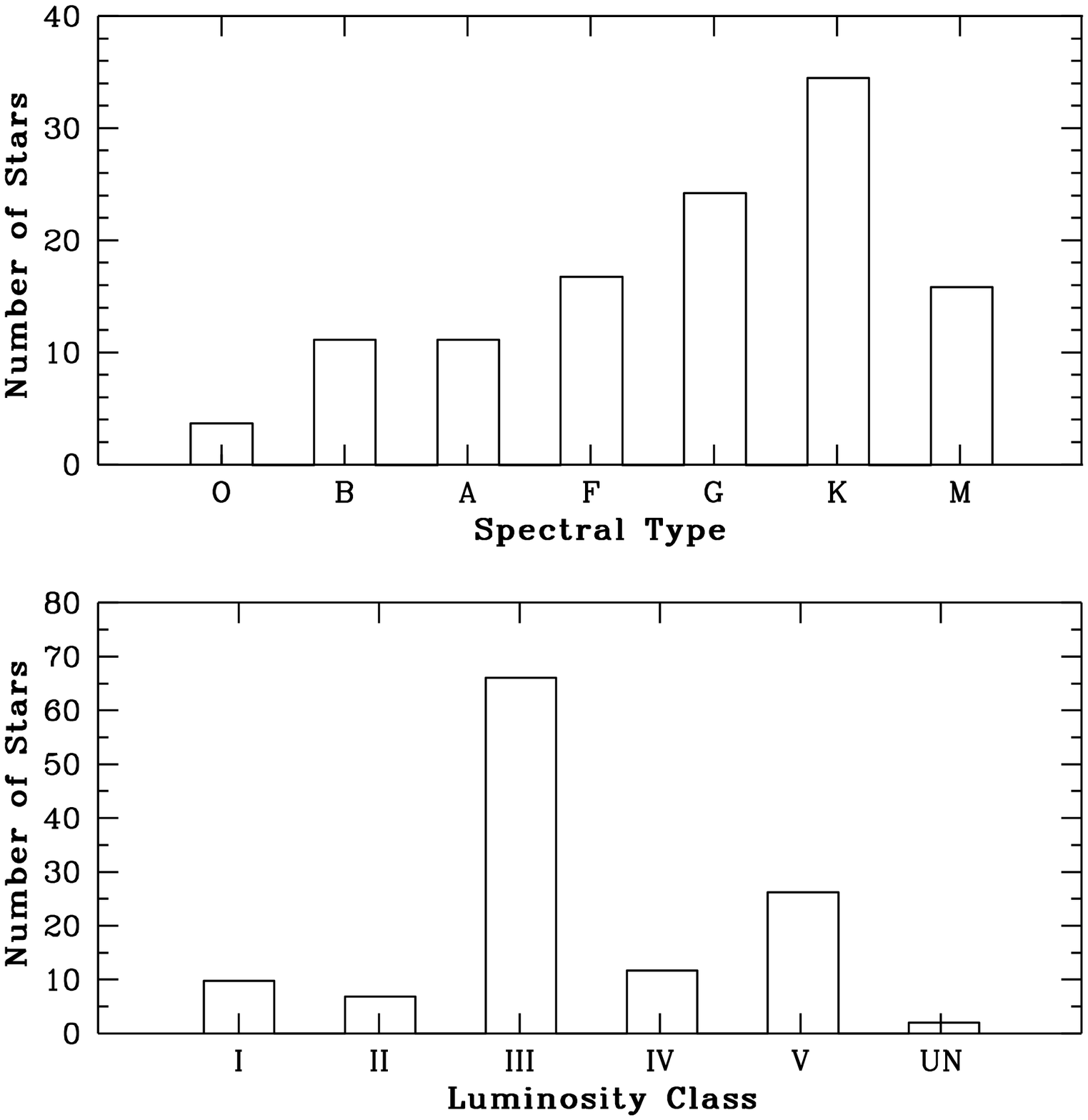,height=18cm,width=15cm}
\caption{Distribution of stars in the database by spectral type and
luminosity class}
\end{figure}

\begin{figure}
\epsfig{file=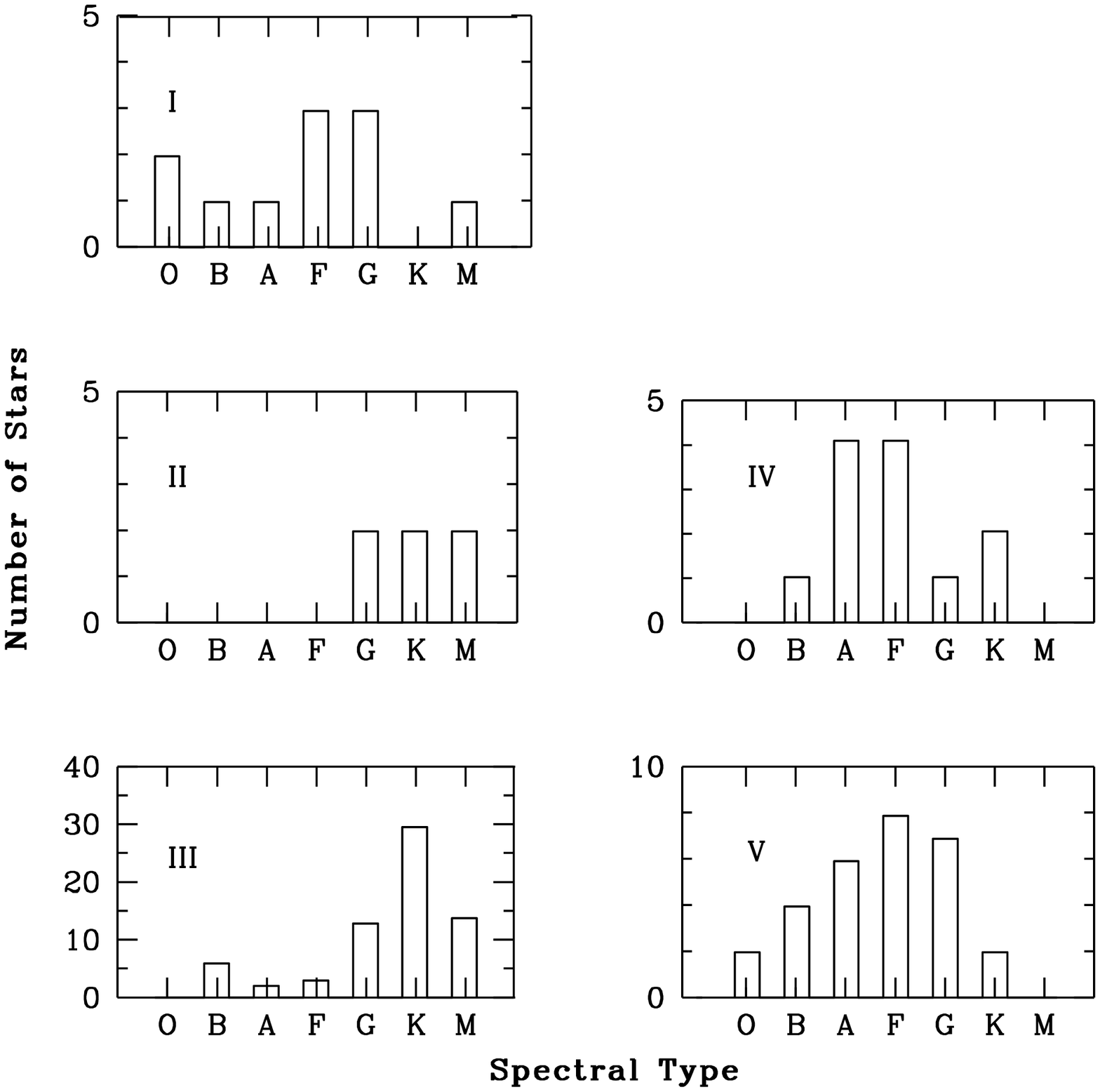,height=18cm,width=15cm}
\caption{Distribution of stars in the database by spectral type per
luminosity class}
\end{figure}

\begin{figure}
\epsfig{file=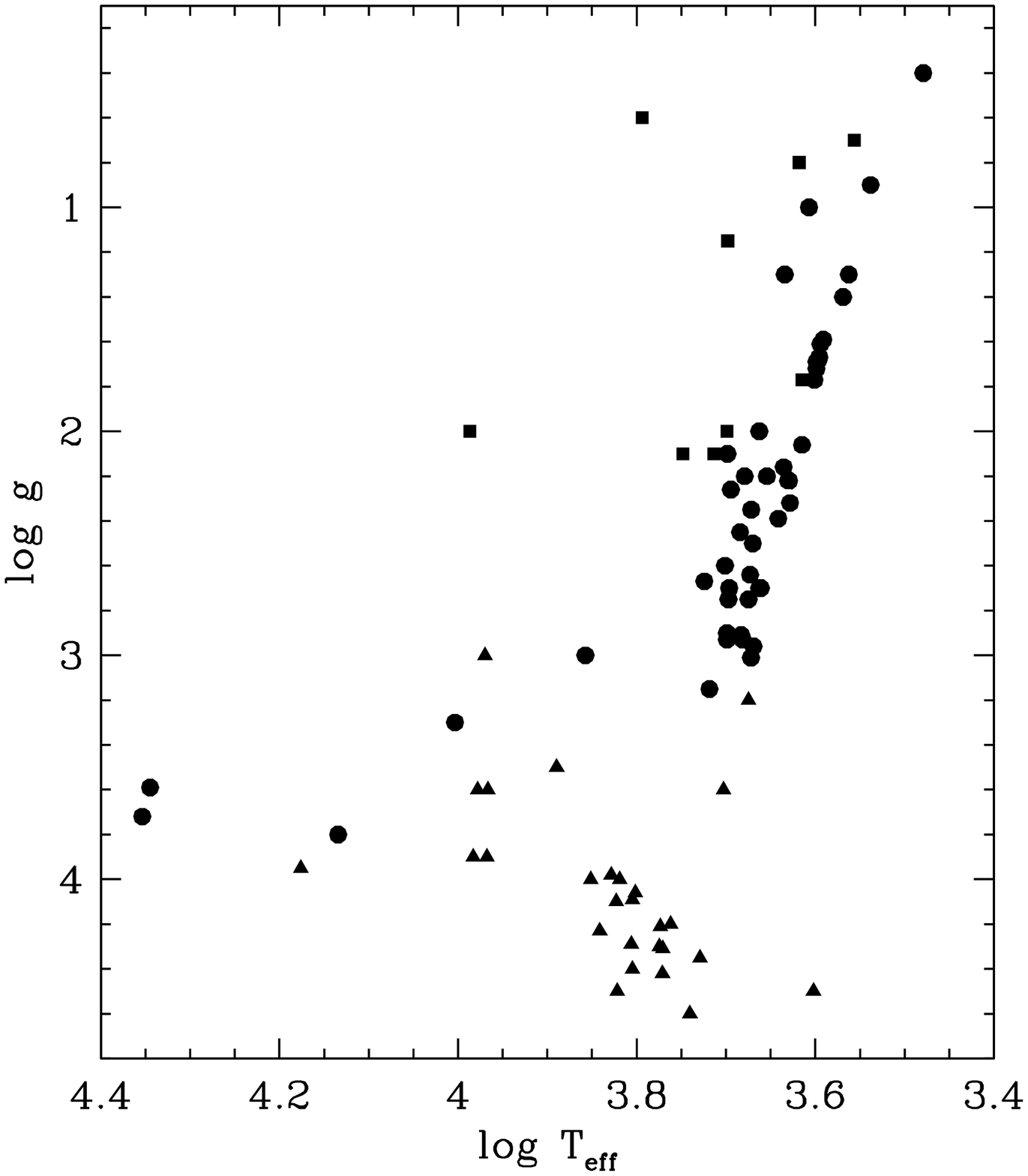,height=18cm,width=15cm} \caption{ Surface
gravity ({\it log g}) vs. effective temperature (T$_{eff})$ for
supergiants  (filled squares), giants (filled circles) and dwarfs
(filled triangles). }
\end{figure}

\begin{table}
\caption{NIR magnitudes of program stars}
\begin{tabular}{ccccccl} \hline
HD & J$_{mag}$ & H$_{mag}$ &
K$_{mag}$ & L$_{mag}$ &
M$_{mag}$ & Reference\\
(1) & (2) & (3) & (4) & (5) & (6) & (7)\\ \hline

HD007927 & 3.75 & 3.54 & 3.19 &      &        & 2003yCat2246...   (Cutri) \\
HD008538 &      & 2.30 &      &      &        & 1998ApJ....508...397 (Meyer) \\
HD010307 & 4.0  & 3.70 & 3.58 &      &        & 2003yCat2246...   (Cutri) \\
HD011353 & 1.92 & 1.36 & 1.22 & 1.60 &        & 1983A\&AS....51...489 (Koornneef) \\
HD023475 & 0.57 &-0.42 &-0.65 &      &        & 2003yCat2246...   (Cutri) \\
HD025204 & 3.66 & 3.67 &3.66 &  3.65 &  3.71  & 1983A\&AS....51...489 (Koornneef) \\
HD026846 & 2.97 & 2.39 &2.27 &  2.19 &        & 1983A\&AS....51...489 (Koornneef) \\
HD030652 & 2.37 & 2.15 &2.08 &  2.09 &        & 1983A\&AS....51...489 (Koornneef) \\
HD030836 & 4.03 & 4.10 &4.15 & 4.20  &        & 1990MNRAS...242...1 (Carter) \\
HD035468 & 2.17 & 2.28 &2.32 &  2.34 &  2.36  & 1983A\&AS....51...489 (Koornneef) \\
HD035497 & 1.96 & 2.02 &2.02 &  2.03 &  2.11  & 1983A\&AS....51...489 (Koornneef) \\
HD036673 & 2.05 & 1.92 &1.86 &  1.81 &        & 1983A\&AS....51...489 (Koornneef) \\
HD037128 & 2.19 & 2.40 &2.27 &       &        & 2003yCat2246...   (Cutri) \\
HD037742 & 2.21 & 2.27 &2.32 &  2.31 &        & 1983A\&AS....51...489 (Koornneef) \\
HD038393 & 2.70 & 2.47 &2.41 &  2.38 & 2.42   & 1983A\&AS....51...489 (Koornneef)  \\
HD038858 & 4.82 & 4.50 &4.44 &       &        & 1991A\&AS....91...409 (Bouchet) \\
HD040136 & 3.10 & 2.94 &2.90 &  2.87 &        & 1983A\&AS....51...489 (Koornneef) \\
HD043232 & 1.84 & 1.19 &1.02 &  0.94 &        & 1983A\&AS....51...489 (Koornneef) \\
HD047105 & 1.87 & 1.86 &1.86 & 1.84 & 1.83    & 1991A\&AS...91...409 (Bouchet) \\
HD047839 & 5.2  & 5.32 & 5.34&       &        & 2003yCat2246...   (Cutri) \\
HD048329 & 0.89 & 0.23 & 0.13&       &        & 2003yCat2246...   (Cutri) \\
HD049331 & 1.63 & 0.86 & 0.56&       &        & 2003yCat2246...   (Cutri) \\
HD054605 & 0.77 & 0.51 &0.41 &  0.32 &  0.28  & 1983A\&AS....51...489 (Koornneef) \\
HD054810 & 3.18 & 2.64 &2.53 &  2.47 & 2.58   & 1983A\&AS....51...489 (Koornneef) \\
HD056537 & 3.54 & 3.50 &3.54 &       &        & 2003yCat2246...   (Cutri) \\
HD058715 & 1.83 & 1.07 &0.90 &  0.77 & 2.94   & 1983A\&AS....51...489 (Koornneef) \\
HD060414 & 1.25 & 0.38 &0.09 &  -0.09&  0.17  & 1983A\&AS....51...489 (Koornneef) \\
HD061421 & -0.39&-0.59 &-0.63 &-0.70 &-0.70   & 1991A\&AS...91...409 (Bouchet) \\
HD061935 & 2.28 & 1.77 &1.62 &  1.57 &1.70    & 1983A\&AS....51...489 (Koornneef) \\
HD062345 & 2.05 & 1.64 & 1.52 &      &        & 2003yCat2246...   (Cutri) \\
HD062576 & 1.74 & 0.96 &0.75 &  0.63 &0.88    & 1983A\&AS....51...489 (Koornneef) \\
HD062721 & 2.07 & 2.27 &1.24 &       &        & 2003yCat2246...   (Cutri) \\
HD063700 & 1.52 & 1.03 &0.89 &  0.81 &        & 1983A\&AS....51...489 (Koornneef) \\
HD066811 & 2.79 & 2.96 &2.97 &       &        & 2003yCat2246...   (Cutri) \\
HD067228 & 4.13 & 3.91 &3.83 &  3.79 &  3.92  & 1983A\&AS....51...489 (Koornneef) \\
HD068312 & 3.79 & 3.23 &3.15 &       &        & 2003yCat2246...   (Cutri) \\
HD070272 & 1.26 & 0.45 &0.38 &       &        & 2003yCat2246...   (Cutri) \\
HD071369 &      &      &     &       &        & 1997ApJ....111...445 (Wallace) \\
HD072094 & 2.45 & 1.64 &1.43 &  1.26 &  1.57  & 1994A\&AS....105...311 (Fluks) \\
HD074918 & 2.80 & 2.33 &2.23 &  2.17 &        & 1983A\&AS....51...489 (Koornneef) \\
HD076943 &      &      &     &       &        & 2004ApJS...152..251 (INDO-US) \\
HD077912 & 2.84 & 2.46 & 2.4 &       &        & 2003yCat2246...   (Cutri) \\
HD080874 & 1.69 & 0.84 & 0.60& 0.38  & 0.67   & 1991A\&AS...91...409 (Bouchet)\\
HD081797 & -0.36& -1.04& -1.21&-1.33 & -1.6   & 1983A\&AS...51...489 (Koornneef)\\
HD082328 &      &      &      &      &        & 2004ApJS...152..251 (INDO-US) \\
HD084748 & -0-.72&-1.75&     &       &        & 2003yCat2246...   (Cutri) \\
HD085444 & 2.59 & 2.13 &2.02 &  1.97 & 2.06   & 1983A\&AS....51...489 (Koornneef) \\
\hline
\end{tabular}
\end{table}

\begin{table}
\setcounter{table}{3}
%\caption{table3 contd..\\}
\begin{tabular}{ccccccl} \hline
HD & J$_{mag}$ & H$_{mag}$ &
K$_{mag}$ & L$_{mag}$ &
M$_{mag}$ & Reference\\
(1) & (2) & (3) & (4) & (5) & (6) & (7)\\ \hline

HD085951 & 2.01 & 1.22 &1.01 &  0.85 &  1.18  & 1994A\&AS....105...311 (Fluks) \\
HD086663 & 1.54 & 0.72 &0.50 &  0.34 &  0.66  & 1994A\&AS....105...311 (Fluks) \\
HD087737 & 3.50 & 3.50 &3.30 &       &        & 2003yCat2246...   (Cutri) \\
HD088230 & 3.89 & 3.30 &2.96 &       &        & 2003yCat2246...   (Cutri) \\
HD088284 & 1.99 & 1.51 &1.40 &  1.34 &  1.48  & 1983A\&AS....51...489 (Koornneef) \\
HD089025 & 2.7  & 2.62 &2.63 &       &        & 2003yCat2246...   (Cutri) \\
HD089021 & 3.44 & 3.46 &3.42 &       &        & 2003yCat2246...   (Cutri) \\
HD089449 & 4.04 & 3.94 &4.02 &       &        & 2003yCat2246...   (Cutri) \\
HD089490 & 4.86 & 4.45 &4.32 &       &        & 2003yCat2246...   (Cutri) \\
HD089758 & -0.11&-0.91 &-1.01&       &        & 2003yCat2246...   (Cutri) \\
HD090254 & 2.45 & 1.59 &1.36 &  1.20 &  1.48  & 1994A\&AS....105...311 (Fluks) \\
HD090432 & 1.31 & 0.56 &0.38 &  0.26 &        & 1983A\&AS....51...489 (Koornneef)\\
HD090610 & 1.81 & 1.07 &0.91 &  0.77 &  1.00  & 1994A\&AS....105...311 (Fluks) \\
HD092125 & 3.30 & 2.93 &2.83 &       &        & 2003yCat2246...   (Cutri) \\
HD092588 & 4.69 & 4.54 &4.15 &       &        & 2003yCat2246...   (Cutri) \\
HD093813 & 1.07 & 0.42 &0.27 &  0.17 &        & 1983A\&AS....51...489 (Koornneef) \\
HD094264 & 2.05 & 1.43 &1.55 &       &        & 2003yCat2246...   (Cutri) \\
HD094481 & 4.24 & 3.76 &3.75 &       &        & 2003yCat2246...   (Cutri) \\
HD095418 & 2.27 & 2.36 &2.29 &       &        & 2003yCat2246...   (Cutri) \\
HD097603 & 2.32 & 2.27 &2.27 &  2.29 &        & 1983A\&AS....51...489 (Koornneef) \\
HD097778 & 1.01 & 0.16 &-0.07 &       &       & 2003yCat2246...   (Cutri) \\
HD098231 &      &      &     &       &        & 2004ApJS...152..251 (INDO-US) \\
HD099028 &      &      &     &       &        & 2004ApJS...152..251 (INDO-US) \\
HD099167 & 1.93 & 1.19 &1.01 &       &        & 2003yCat2246...   (Cutri) \\
HD100920 & 2.78 & 2.27 &2.18 &       &        & 2003yCat2246...   (Cutri) \\
HD101501 & 3.99 & 3.65 &3.59 &       &        & 2003yCat2246...   (Cutri) \\
HD102212 & 1.08 & 0.23 &0.03 &-0.09  &        & 1991A\&AS...409...424 (Bouchet)\\
HD105707 & 0.94 & 0.31 &0.14 &  0.03 & 0.19   & 1983A\&AS....51...489 (Koornneef) \\
HD106625 & 2.79 & 2.83 &2.82 &  2.76 & 2.80   & 1983A\&AS....51...489 (Koornneef) \\
HD107259 & 3.81 & 3.78 &3.77 &  3.76 &        & 1990MNRAS....242...1 (Carter) \\
HD107328 & 2.95 & 2.32 &2.19 &  2.09 & 2.30   & 1983A\&AS....51...489 (Koornneef) \\
HD109358 & 3.21 & 2.90 &2.85 &       &        & 2003yCat2246...   (Cutri) \\
HD109379 & 1.24 & 0.81 &0.70 &  0.64 & 0.72   & 1983A\&AS....51...489 (Koornneef) \\
HD110379 & 2.07 & 1.90 &1.86 &  1.84 & 1.92   & 1983A\&AS....51...489 (Koornneef) \\
HD111812 & 3.73 & 3.46 &3.36 &  3.29 &  3.34  & 1983A\&AS....51...489 (Koornneef) \\
HD112142 & 1.27 & 0.40 &0.17 &       &        & 2003yCat2246...   (Cutri) \\
HD112300 & -0.11&-1.01 &-1.19&       &        & 2003yCat2246...   (Cutri) \\
HD113139 & 4.32 &4.16  & 3.95&       &        & 2003yCat2246...   (Cutri) \\
HD113226 & 1.25 &0.73  & 0.66&       &        & 2003yCat2246...   (Cutri) \\
HD113847 & 3.76 &3.15  & 2.90&       &        & 2003yCat2246...   (Cutri) \\
HD113996 & 2.37 &1.63  & 1.49&       &        & 2003yCat2246...   (Cutri) \\
HD114330 &      &      &     &       &        & 1997ApJ....111...445 (Wallace) \\
HD114961 &-0.36 &-1.61 &     &       &        & 2004ApJS...152..251 (INDO-US) \\
HD115604 &4.06  &4.02  &4.01 &       &        & 2003yCat2246...   (Cutri) \\
HD115659 &1.48  &1.03  &0.94 &0.89   &        &1991A\&AS...409...424 (Bouchet) \\
HD115892 & 2.73 & 2.74 &2.73 &  2.70 &        & 1983A\&AS....51...489 (Koornneef) \\
HD116656 &      &      &     &       &        & 1997ApJ....111...445 (Wallace) \\
HD116658 & 1.53 & 1.64 &1.68 &  1.72 & 1.76   & 1983A\&AS....51...489 (Koornneef) \\
\hline
\end{tabular}
\end{table}

\begin{table}
\setcounter{table}{3}
%\caption{table3 contd..\\}
\begin{tabular}{ccccccl} \hline
HD & J$_{mag}$ & H$_{mag}$ &
K$_{mag}$ & L$_{mag}$ &
M$_{mag}$ & Reference\\
(1) & (2) & (3) & (4) & (5) & (6) & (7)\\ \hline
HD116870 & 2.62 & 1.81 &1.61 &  1.47 &  1.73 & 1994A\&AS....105...311 (Fluks) \\
HD120052 & 1.88 & 1.02 &0.73 &       &       & 2003yCat2246...   (Cutri) \\
HD120315 & 2.23 & 2.41 &2.27 &       &       & 2003yCat2246...   (Cutri) \\
HD121299 & 3.66 & 2.98 &2.86 &       &       & 2003yCat2246...   (Cutri) \\
HD123123 & 1.42 & 0.84 &0.72 & 0.62  &       &1983A\&AS...51...489 (Koornneef) \\
HD123139 & 0.42 &-0.09 &-0.21&  -0.31&  -0.2 1& 1983A\&AS....51...489 (Koornneef) \\
HD123299 & 3.43 &3.63  &3.64 &       &        & 2003yCat2246...   (Cutri) \\
HD123657 & 1.01 &-0.01 &-0.23&       &        & 2003yCat2246...   (Cutri) \\
HD123934 & 1.65 &0.81  &-0.60&       &        & 2003yCat2246...   (Cutri) \\
HD124294 & 1.89 & 1.18 &1.03 &  0.94 &        & 1983A\&AS....51...489 (Koornneef) \\
HD126661 & 5.16 & 4.94 &4.87 &       &        & 2003yCat2246...   (Cutri) \\
HD127665 & 1.50 & 0.76 &0.76 &       &        & 2003yCat2246...   (Cutri) \\
HD129116 & 4.38 & 4.46 &4.52 &  4.56 &        & 1990MNRAS....242...1 (Carter) \\
HD129502 & 3.12 & 2.94 &2.89 &  2.83 &        & 1983A\&AS....51...489 (Koornneef) \\
HD130841 & 2.52 & 2.45 &2.42 &  2.40 &  2.42  & 1983A\&AS....51...489 (Koornneef) \\
HD130952 & 3.50 & 2.90 &2.80 &       &        & 2003yCat2246...   (Cutri) \\
HD131156 &      &      &     &       &        & 2004ApJS...152..251 (INDO-US) \\
HD131918 & 3.01 &2.28  &2.09 &       &        & 2003yCat2246...   (Cutri) \\
HD134083 & 4.25 &4.01  &3.86 &       &        & 2003yCat2246...   (Cutri) \\
HD135722 & 1.66 &0.99  &1.22 &       &        & 2003yCat2246...   (Cutri) \\
HD136512 & 3.65 &3.17  &2.93 &       &        & 2003yCat2246...   (Cutri) \\
HD138716 & 2.86 &2.34  &2.24 &       &        & 2003yCat2246...   (Cutri) \\
HD138905 & 2.24 &1.67  &1.55 &1.47   &        & 1991A\&AS...409...424 (Bouchet)\\
HD141004 & 3.36 & 3.05 &2.99 &       &        & 1991A\&AS....91...409 (Bouchet) \\
HD141714 & 3.21 & 2.80 &2.67 &       &        & 2003yCat2246...   (Cutri) \\
HD141850 & 2.05 & 1.23 &0.69 &  -0.08&  -0.10 & 1994A\&AS....105...311 (Fluks) \\
HD145328 & 2.99 & 2.50 &2.34 &       &        & 2003yCat2246...   (Cutri) \\
HD147165 & 2.49 & 2.44 &2.42 &  2.42 &  2.43  & 1983A\&AS....51...489 (Koornneef) \\
HD147394 &      &      &     &       &        & 2004ApJS...152..251 (INDO-US) \\
HD148513 & 2.95 &2.16  &2.04 &1.92   &        & 1990MNRAS....242...1 (Carter) \\
HD149752 & 8.65 &8.55  & 8.55 &      &        & 2003yCat2246...   (Cutri) \\\\
\hline
\end{tabular}
\end{table}
\begin{figure}
\epsfig{file=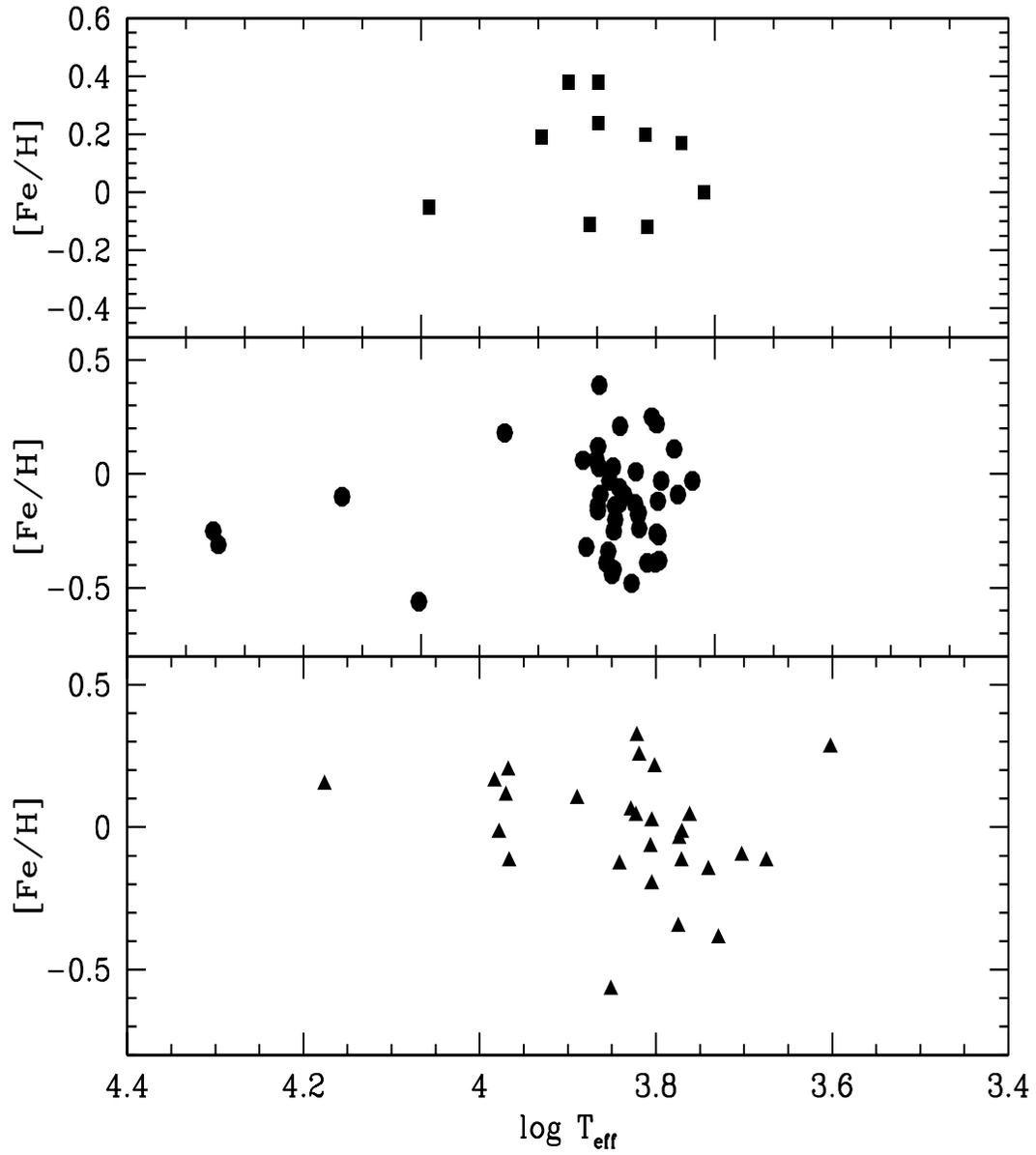,height=18cm,width=15cm} \caption{Metallicity
[Fe/H] vs. effective temperature (T$_{eff}$) for supergiants (filled
squares), giants (filled circles) and dwarfs (filled triangles)
({\it from top to bottom}).}
\end{figure}

\begin{figure}
\epsfig{file=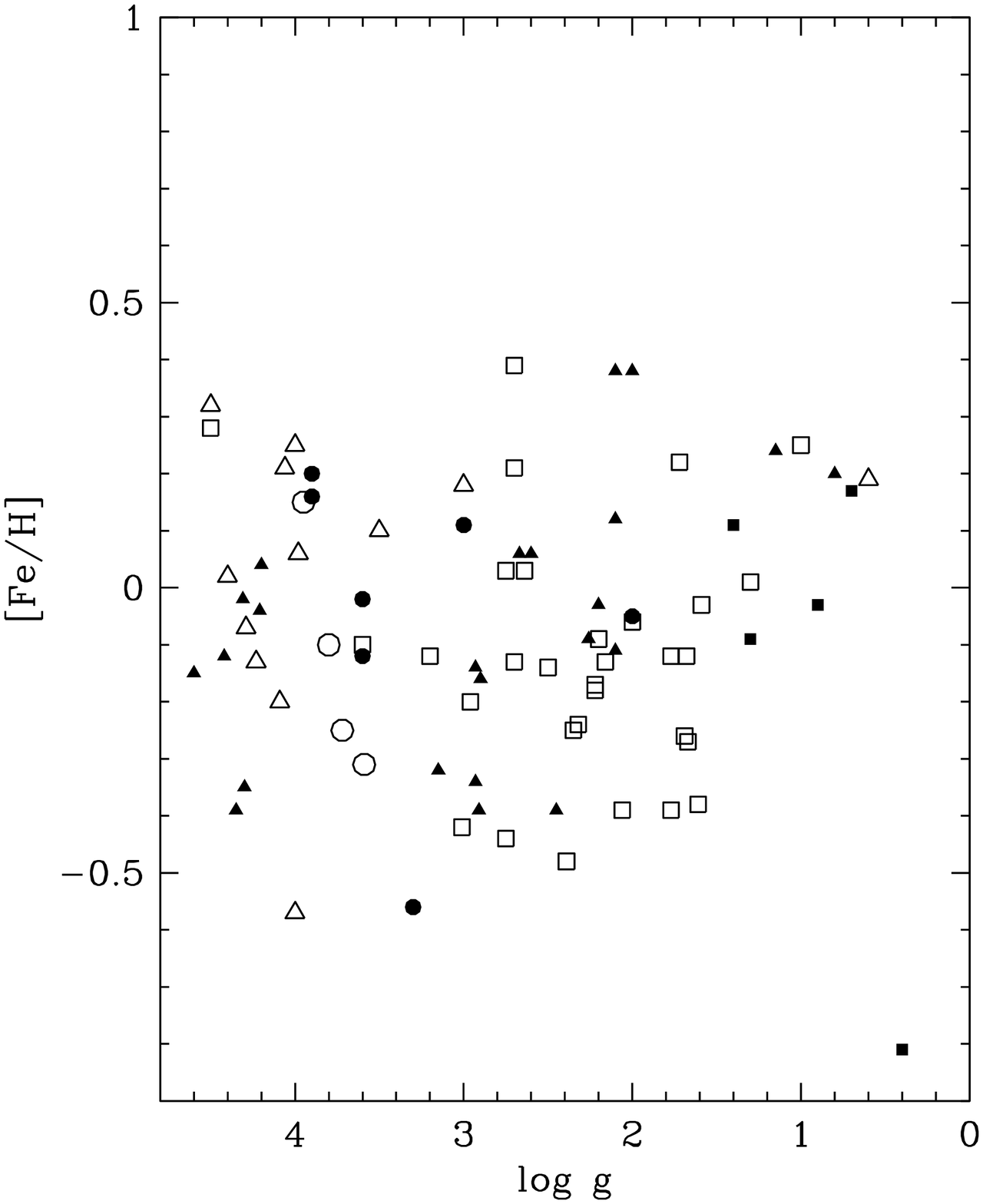,height=18cm,width=15cm} \caption{Metallicity
[Fe/H] vs. surface gravity ({\it log g}) for spectral types B (open
circles), A (filled circles) F (open triangles), G (filled
triangles), K (open squares) and M (filled squares).}
\end{figure}

\section{DATA REDUCTION AND CALIBRATION}

 The near infrared spectral data reduction is similar to that of optical
data reduction with minor differences. The presence of strong
telluric emission lines and varying atmospheric transmission due to
changing water vapour content necessitates observation of standard
star spectra at similar airmass soon after the program star
observation. The whole process of near infrared long slit spectra
reduction can be separated into a few major steps, viz., (i)
pre-processing (ii) spectrum extraction (iii) wavelength calibration
(iv) atmospheric transmission and instrument response determination
using standard star data (v) continuum fitting and (vi) radial
velocity correction. We have used standard tasks available in
software package IRAF \footnote {IRAF is distributed by National
Optical Astronomy Observatories, which are operated by the
Association of Universities for Research in Astronomy, Inc., under
cooperative agreement with the National Science Foundation.} for
data reduction. As discussed in \S2 we have two sets of frames at
two different locations of the detector. The availability of these
two sets of spectra is utilized to remove the dark counts and the
large sky background at near infrared wavelengths. This is
accomplished by taking the difference of  spectra obtained at two
different locations on the detector. As there is no autoguider on
the telescope, the frames with maximum counts in two positions are
selected for data reduction. We thus have two difference frames for
extraction of the spectrum. The detail of the each task and its
significance in the data reduction is discussed in paper I. The
important aspect of the data reduction is to perform the wavelength
calibration. The telluric absorption features at 11354 \AA~ and
12684 \AA~ in both programme and standard stars were used for
wavelength calibration. The IRAF task {\it{identify}} is used for
this purpose. The IRAF task {\it{refspec}} is used to specify the
appropriate wavelength calibrated spectrum for the stellar spectra
extracted through {\it{apall}} task. The IRAF task {\it{dispcor}}
was used to set the wavelength calibration for the stellar spectra.
The effects like atmospheric transmission effects and the instrument
effects (filter transmission and wavelength dependence of detector
quantum efficiency) can be removed by taking the ratio of the
program star spectrum with that of a standard star spectrum observed
under similar conditions. We have selected bright A and late B type
with T$_{eff} \approx $ 10000 K because at this temperature only
neutral hydrogen lines will be present and no metallic lines will
present in the NIR spectral region. Table 2 lists standard stars
used for the purpose of taking ratios. The stellar absorption
feature due to hydrogen namely the Paschen $\beta$ line was removed
before taking the ratio. The program star flux is divided by the
corresponding standard star flux and in this process the modulation
due to telluric features, atmospheric extinction and instrumental
effects cancels out. The resultant spectrum from this division is
multiplied with a corresponding blackbody flux distribution at the
temperature corresponding to the standard star. It may be noted that
unlike many of the spectral libraries published earlier, the spectra
presented here have been continuum shape corrected to their
respective
 effective temperatures. The list of 126 programme stars selected in this library is
shown in table 4. In this, the first and second column
contain the star ID, columns (3) and (4) contain the right ascension
(J2000.0) and declination (J2000.0) respectively and column (5) gives the apparent $V$ magnitude.
The column (6) gives the corresponding standard star ID used for data reduction.

\section{SPECTRAL LIBRARY}

The NIR J band spectral library of 126 stars is available in the format
of reduced ASCII tables with wavelength versus flux at a spectral resolution
of 1000 at 5 \AA~ binning. The main goal of this paper is to make this library
available for variety of investigators working in the NIR region.
Thus, the complete library can be downloaded from the website:

http://vo.iucaa.ernet.in/$\sim$voi/NIR\_Header.html

The essential information of each star in the database is summarized
in Tables 3 and 4 as observational parameters and in Table 5 as
physical parameters. The contents of Table 3 has been mentioned in
section \S3 and content of table 4 has been mentioned in section
\S4. Table 5 contains the star ID in the first column. Columns (2),
(3) and (4) give spectral type, luminosity class and effective
temperature respectively. Columns (5) and (6) give the $\it{log g}$
and [Fe/H] values respectively. The last column gives the references
from which the physical parameters have been obtained.

    Fig. 6, shows spectra of seven supergiant stars, covering a large range of
MK spectral type, and thus illustrates the basic dependence of
spectral features on spectral types. Fig. 7 shows spectra of six
giant stars again covering different spectral types. Similarly Fig.
8 shows a series of six dwarf stars. All of these plots illustrate
the change in basic features with the temperature, gravity and
metallicity. We also attempt to show the quality of spectra by
comparing some selected spectra with the already published J band
library by Wallace et al. (2000).

Following paragraphs describe the procedure that we have followed for
comparing the GIRT and Wallace data.

The block diagram in Fig. 9 depicts the steps carried out
on both libraries. There are two steps performed on the library by
Wallace et al. (2000).

(i) conversion of wavenumber vs. relative flux to wavelength vs. relative flux.

(ii) fitting a continuum to respective T$_{eff}$ of each star and
lagrange fitting for binning at 5 \AA~ steps.

These steps were performed by writing a common algorithm which could
run uniformly on the Wallace et al. (2000) library.
The T$_{eff}$ values were taken from 'Astronomical Hand Book' by K. R. Lang
and the black body spectra were generated by IRAF {\it{mk1dspec}} task
for J band region.

\begin{figure}
\epsfig{file=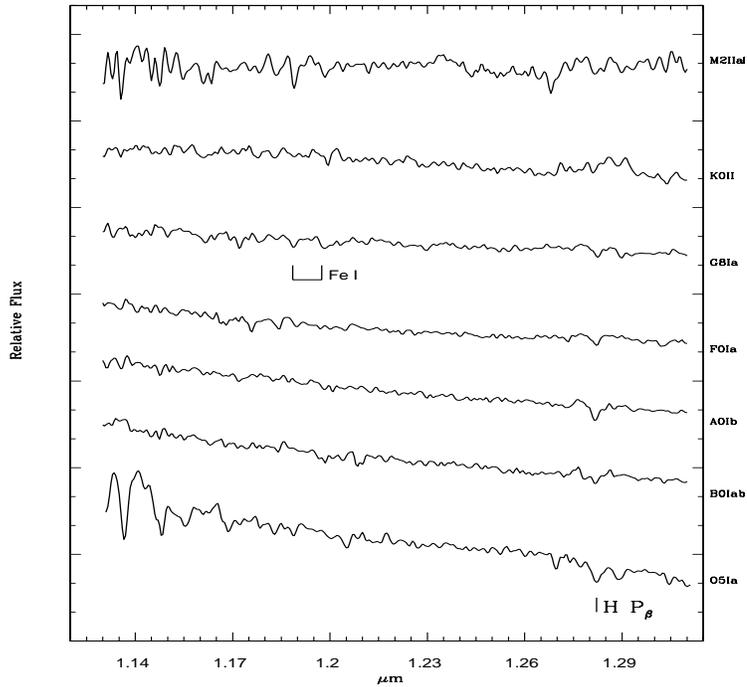,height=10cm,width=10cm} \caption{Spectra of
seven supergiant stars, covering a large range of MK spectral type,
are plotted to illustrate the basic dependence of spectral features
on spectral type. The stars plotted are (top to bottom) HR1155,
HR4255, HR2473, HR382, HR3975, HR1903 and HR3165. The spectral types
are listed on the side.}
\end{figure}

\begin{figure}
\epsfig{file=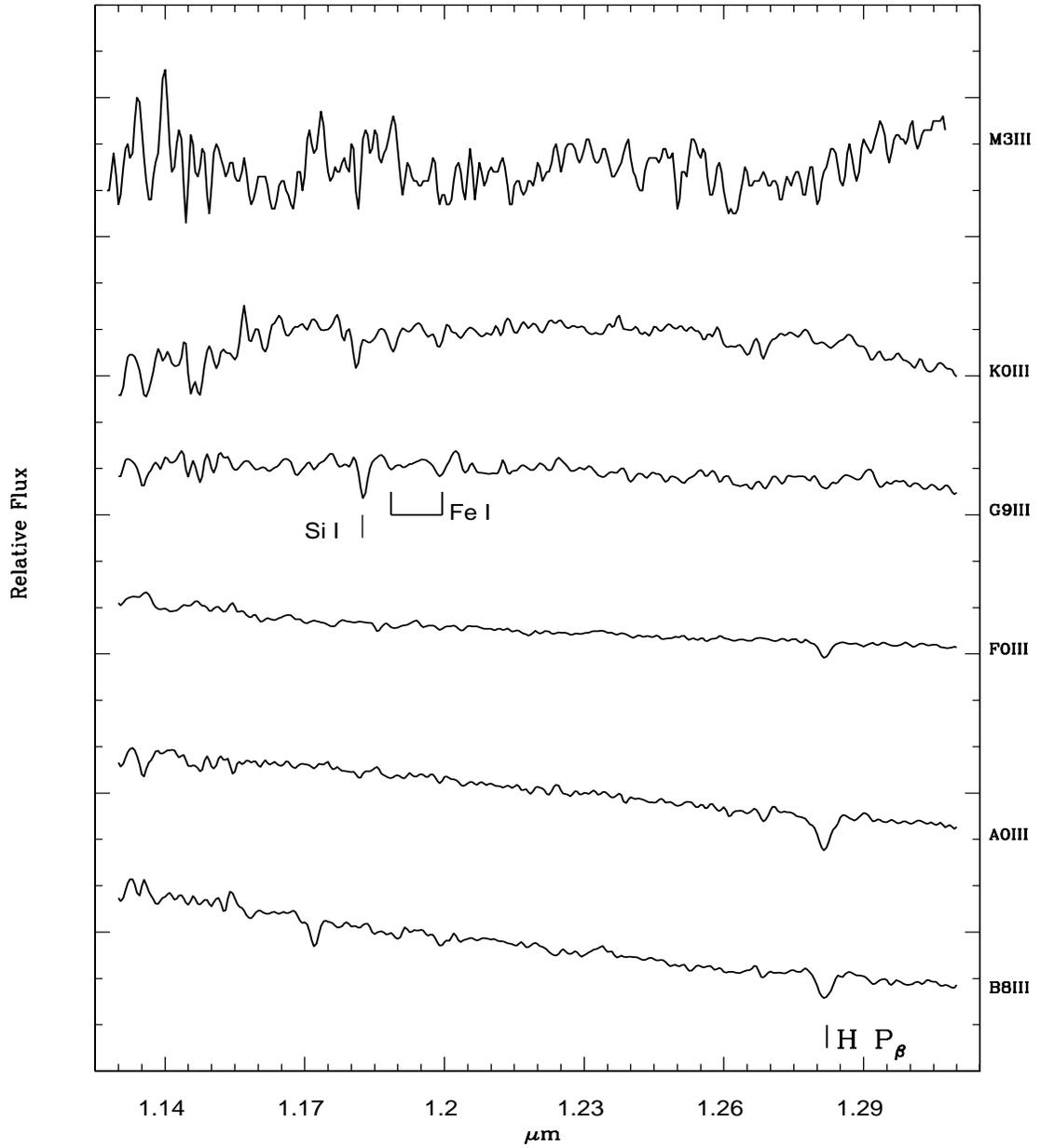,height=18cm,width=15cm} \caption{Spectra of six
giant stars are plotted to illustrate the basic dependence of
spectral features on spectral type. The stars are (top to bottom)
HR4517, HR4232, HR2970, HR4031, HR5291 and HR4662. The spectral
types are listed on the side.}
\end{figure}

\begin{figure}
\epsfig{file=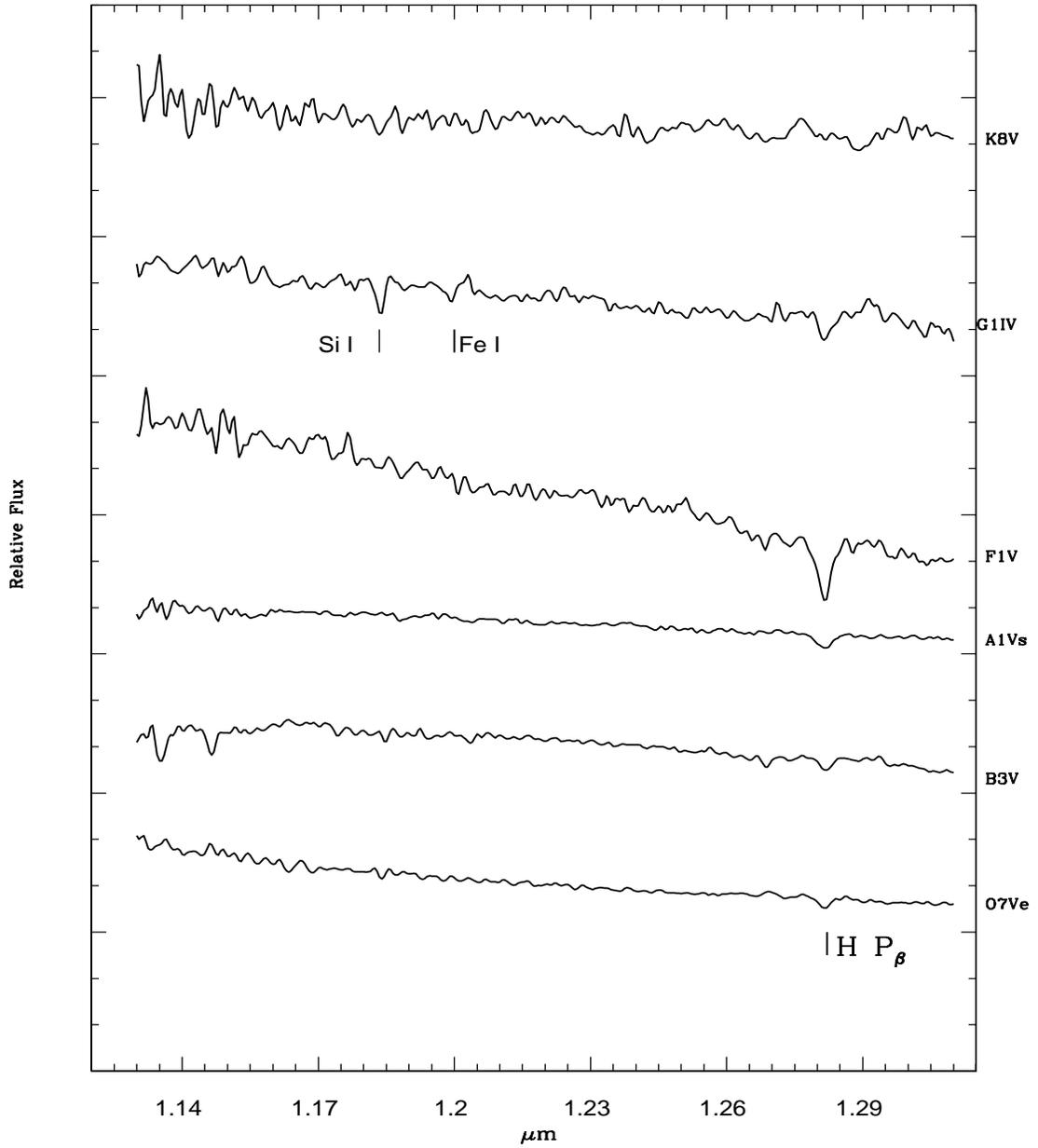,height=18cm,width=15cm} \caption{Spectra of six
dwarf stars (top to bottom) HD88230, HR3176, HR2085, HR4963, HR5471
and HR2456. The spectral types are listed on the side.}
\end{figure}

\begin{figure}
\epsfig{file=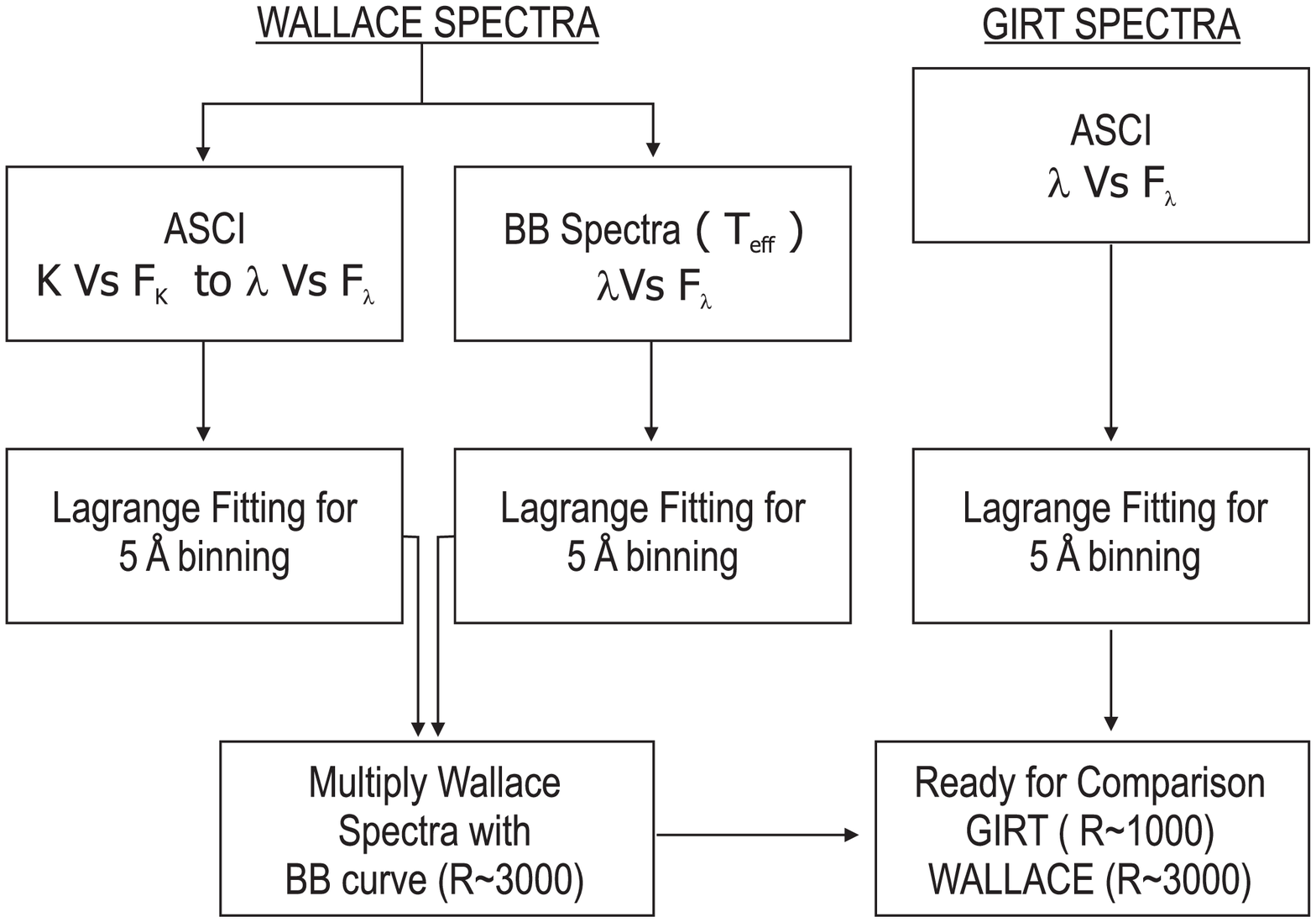,height=18cm,width=15cm}
\caption{Block diagram illustrating the steps involved in comparison of
GIRT and Wallace et al. (2000) libraries}
\end{figure}

\begin{figure}
\epsfig{file=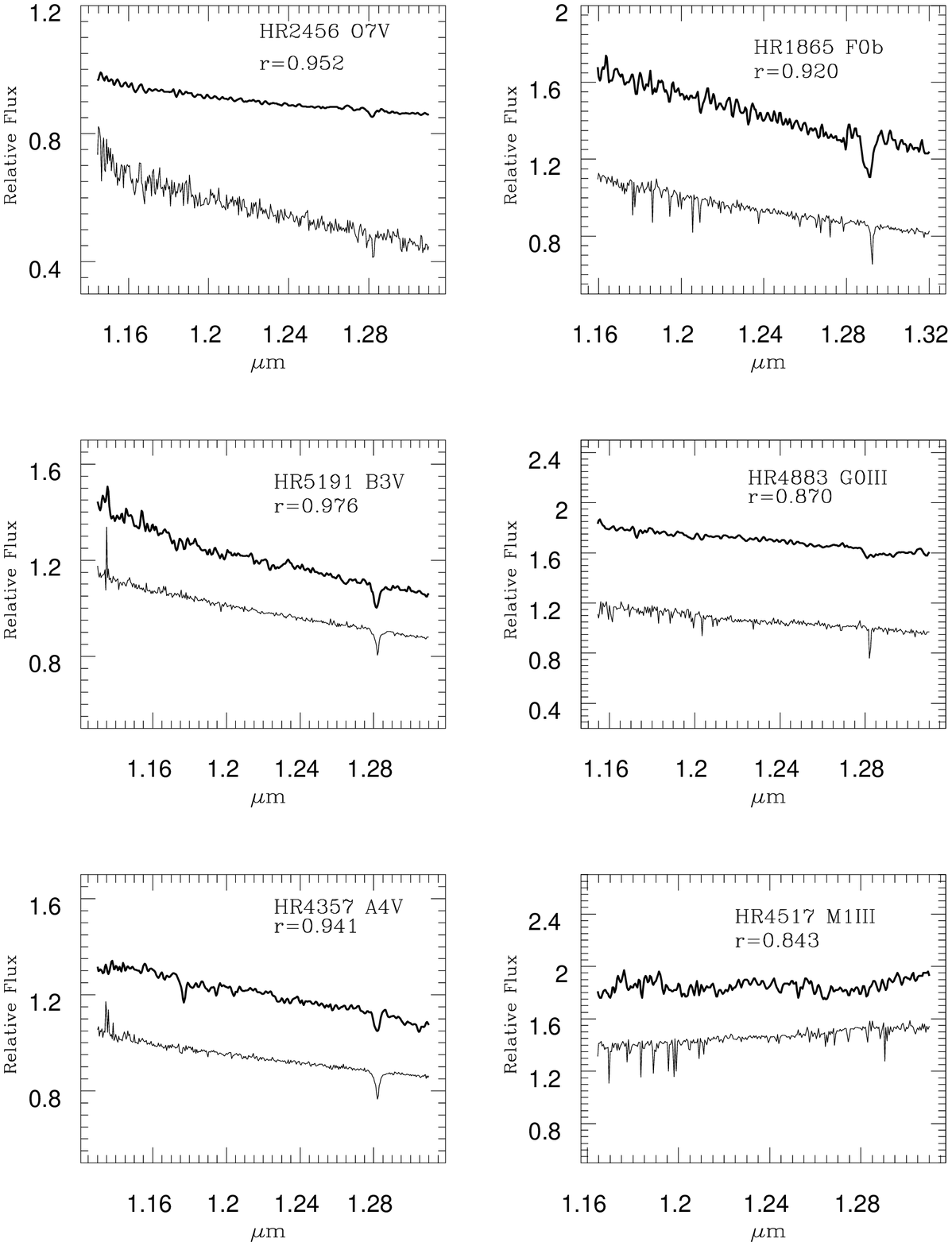,height=18cm,width=15cm} \caption{A selection of
common spectra from Wallace et al. 2000 ({\it thin} lines) and GIRT
({\it thick} lines) libraries. Please note that the two spectra in
each panel have been offset purposely for sake of clarity and the
flux values are relative.}
\end{figure}

Fig. 10 shows a sample of some of the common stars in GIRT and
Wallace et al. library with good matching of the spectral features
as evident from the correlation coefficient {\it{r}} values. This
plot covers most of the main spectral types. It may be noted that
the resolution of both the spectra is not same viz. GIRT $\sim$ 1000
and Wallace $\sim$ 3000.

\begin{table}
\caption{Observational Parameters (from SIMBAD) of program stars}
\begin{tabular}{cccccc} \hline
HD & HR & $\alpha$(J2000.0) &
$\delta$(J2000.0) &
V$_{mag}$ & Standard Star \\
& & & & & \\
(1) & (2) & (3) & (4) & (5) & (6) \\ \hline
HD007927 & HR382  & 01 20 04.91 & +58 13 53.79  &  5.01 & HR3314  \\
HD008538 & HR403  & 01 25 48.95 & +60 14 07.01  &  2.68 & HR3314  \\
HD010307 & HR483  & 01 41 47.14 & +42 36 48.12  &  4.90 & HR3314  \\
HD011353 & HR539  & 01 51 27.63 & -10 20 06.13  &  3.73 & HR3314  \\
HD023475 & HR1155 & 03 49 31.28 & +65 31 33.50  &  4.47 & HR2421  \\
HD025204 & HR1239 & 04 00 40.81 & +12 29 25.24  &  3.40 & HR2421  \\
HD026846 & HR1318 & 04 14 23.68 & -10 15 22.61  &  4.90 & HR2421  \\
HD030652 & HR1543 & 04 49 50.41 & +06 57 40.59  &  3.19 & HR3314  \\
HD030836 & HR1552 & 04 51 12.36 & +05 36 18.37  &  4.47 & HR2421  \\
HD035468 & HR1790 & 05 25 07.86 & +06 20 58.92  &  1.62 & HR3314  \\
HD03549- & HR1791 & 05 26 17.51 & +28 36 26.82  &  1.68 & HR3314  \\
HD036673 & HR1865 & 05 32 43.81 & -17 49 20.23  &  2.59 & HR2421  \\
HD037128 & HR1903 & 05 36 12.81 & -01 12 06.91  &  1.70 & HR3314  \\
HD037742 & HR1948 & 05 40 45.53 & -01 56 33.50  &  1.70 & HR3314  \\
HD038393 & HR1983 & 05 44 27.79 & -22 26 54.17  &  3.60 & HR2421  \\
HD038858 & HR2007 & 05 48 34.94 & -04 05 40.73  &  5.97 & HR2421  \\
HD040136 & HR2085 & 05 56 24.29 & -14 10 03.72  &  3.71 & HR1412  \\
HD043232 & HR2227 & 06 14 51.33 & -06 14 29.19  &  3.98 & HR3314  \\
HD047105 & HR2421 & 06 37 42.70 & +16 23 57.30  &  1.90 & HR3314  \\
HD047839 & HR2456 & 06 40 58.66 & +09 53 44.71  &  4.66 & HR3314  \\
HD048329 & HR2473 & 06 43 55.92 & +25 07 52.04  &  3.01 & HR2421  \\
HD049331 & HR2508 & 06 47 37.22 & -08 59 54.60  &  5.10 & HR3982  \\
HD054605 & HR2693 & 07 08 23.48 & -26 23 35.51  &  1.84 & HR5793  \\
HD054810 & HR2701 & 07 10 13.68 & -04 14 13.58  &  4.92 & HR3314  \\
HD056537 & HR2763 & 07 18 05.57 & +16 32 25.37  &  3.58 & HR2421  \\
HD058715 & HR2845 & 07 27 09.04 & +08 17 21.53  &  2.88 & HR2421  \\
HD060414 & HR2902 & 07 33 47.96 & -14 31 26.01  &  4.97 & HR1412  \\
HD061421 & HR2943 & 07 39 18.11 & +05 13 29.97  &  0.34 & HR3982  \\
HD061935 & HR2970 & 07 41 14.83 & -09 33 04.07  &  3.93 & HR2421 \\
HD062345 & HR2985 & 07 44 26.85 & +24 23 52.77  &  3.57 & HR2421  \\
HD062576 & HR2993 & 07 43 32.38 & -28 24 39.18  &  4.62 & HR2891  \\
HD062721 & HR3003 & 07 46 07.44 & +18 30 36.15  &  4.88 & HR2421  \\
HD063700 & HR3045 & 07 49 17.65 & -24 51 35.22  &  3.33 & HR3113  \\
HD066811 & HR3165 & 08 03 35.04 & -40 00 11.33  &  2.21 & HR2421  \\
HD067228 & HR3176 & 08 07 45.85 & +21 34 54.53  &  5.30 & HR2421  \\
HD068312 & HR3212 & 08 11 33.00 & -07 46 21.14  &  5.35 & HR5793  \\
HD070272 & HR3275 & 08 22 50.10 & +43 11 17.27  &  4.25 & HR3982  \\
HD071369 & HR3323 & 08 30 15.87 & +60 43 05.40  &  3.37 & HR2421  \\
HD072094 & HR3357 & 08 31 35.70 & +18 05 40.00  &  5.33 & HR5793  \\
HD074918 & HR3484 & 08 46 22.53 & -13 32 51.79  &  4.32 & HR3314  \\
HD076943 & HR3579 & 09 00 38.40 & +41 46 58.00  &  3.90 & HR2891  \\
HD077912 & HR3612 & 09 06 31.80 & +38 27 08.00  &  4.50 & HR3894  \\
HD080874 & HR3718 & 09 21 29.59 & -25 57 55.58  &  4.72 & HR2421  \\
HD081797 & HR3748 & 09 27 35.24 & -08 39 30.96  &  2.00 & HR2421  \\
HD082328 & HR3775 & 09 32 51.43 & +51 40 38.28  &  3.20 & HR3975  \\
HD084748 & HR3882 & 09 47 33.49 & +11 25 43.64  &  6.02 & HR5793  \\
HD085444 & HR3903 & 09 51 28.69 & -14 50 47.77  &  4.11 & HR2421  \\
HD085951 & HR3923 & 09 54 52.20 & -19 00 34.00  &  4.93 & HR5793  \\
\hline
\end{tabular}
\end{table}

\begin{table}
\setcounter{table}{4}
%\caption{table4 contd..\\}
\begin{tabular}{cccccc} \hline
HD & HR & $\alpha$(J2000.0) &
$\delta$(J2000.0) &
V$_{mag}$ & Standard Star \\
& & & & & \\
(1) & (2) & (3) & (4) & (5) & (6) \\ \hline

HD086663 & HR3950 & 10 00 12.80 & +08 02 39.00  &  4.64 & HR3799  \\
HD087737 & HR3975 & 10 07 19.95 & +16 45 45.59  &  3.51 & HR3314  \\
HD088230 &        & 10 11 22.14 & +49 27 15.25  &  6.61 & HR3665  \\
HD088284 & HR3994 & 10 10 35.27 & -12 21 14.69  &  3.61 & HR5793  \\
HD089025 & HR4031 & 10 16 41.41 & +23 25 02.31  &  3.44 & HR2421  \\
HD089021 & HR4033 & 10 17 05.79 & +42 54 51.71  &  3.44 & HR2421  \\
HD089449 & HR4054 & 10 19 44.10 & +19 28 15.00  &  4.70 & HR4259  \\
HD089490 & HR4059 & 10 19 32.20 & -05 06 21.00  &  6.30 & HR4359  \\
HD089758 & HR4069 & 10 22 19.74 & +41 29 58.25  &  3.06 & HR2421  \\
HD090254 & HR4088 & 10 25 15.20 & +08 47 25.00  &  5.59 & HR3799  \\
HD090432 & HR4094 & 10 26 05.42 & -16 50 10.64  &  3.83 & HR1412  \\
HD090610 & HR4104 & 10 27 09.10 & -31 04 04.00  &  4.27 & HR4660  \\
HD092125 & HR4166 & 10 38 43.21 & +31 58 34.45  &  4.68 & HR5793  \\
HD092588 & HR4182 & 10 41 24.62 & -01 44 23.50  &  6.26 & HR4359  \\
HD093813 & HR4232 & 10 49 37.48 & -16 11 37.13  &  3.11 & HR5793  \\
HD094264 & HR4247 & 10 53 18.33 & +34 13 07.30  &  3.03 & HR4554  \\
HD094481 & HR4255 & 10 54 17.77 & -13 45 28.92  &  5.66 & HR5793  \\
HD095418 & HR4295 & 11 01 50.47 & +56 22 56.73  &  2.34 & HR5793  \\
HD097603 & HR4357 & 11 14 06.50 & +20 31 25.38  &  2.56 & HR5793  \\
HD097778 & HR4362 & 11 15 12.22 & +23 05 43.80  &  4.58 & HR2421  \\
HD098231 & HR4375 & 11 18 10.90 & +31 31 44.90  &  4.41 & HR1412  \\
HD099028 & HR4399 & 11 23 55.50 & +10 31 45.00  &  3.90 & HR4259  \\
HD099167 & HR4402 & 11 24 36.62 & -10 51 34.90  &  4.83 & HR4357  \\
HD100920 & HR4471 & 11 36 57.02 & -00 49 26.00  &  4.30 & HR4554  \\
HD101501 & HR4496 & 11 41 03.01 & +34 12 05.88  &  5.32 & HR2421  \\
HD102212 & HR4517 & 11 45 51.55 & +06 31 45.75  &  4.05 & HR2421  \\
HD105707 & HR4630 & 12 10 07.48 & -22 37 11.15  &  3.01 & HR5793  \\
HD106625 & HR4662 & 12 15 48.37 & -17 32 30.94  &  2.59 & HR2421  \\
HD107259 & HR4689 & 12 19 54.35 & -00 40 00.49  &  3.89 & HR5793  \\
HD107328 & HR4695 & 12 20 20.98 & +03 18 45.26  &  2.06 & HR2421  \\
HD109358 & HR4785 & 12 33 47.64 & +41 21 12.00  &  4.26 & HR4660  \\
HD109379 & HR4786 & 12 34 23.23 & -23 23 48.33  &  2.65 & HR5793  \\
HD110379 & HR4825 & 12 41 39.60 & -01 26 57.90  &  3.65 & HR3314  \\
HD111812 & HR4883 & 12 51 41.92 & +27 32 26.56  &  4.93 & HR3982  \\
HD112142 & HR4902 & 12 54 21.16 & -09 32 20.38  &  4.80 & HR5793  \\
HD112300 & HR4910 & 12 55 36.20 & +03 23 50.89  &  3.38 & HR5867  \\
HD113139 & HR4931 & 13 00 43.59 & +56 21 58.81  &  4.93 & HR2421  \\
HD113226 & HR4932 & 13 02 10.59 & +10 57 32.94  &  2.83 & HR3982  \\
HD113847 & HR4945 & 13 05 52.30 & +45 16 07.00  &  5.60 & HR4660  \\
HD113996 & HR4954 & 13 07 10.70 & +27 37 29.00  &  4.80 & HR5867  \\
HD114330 & HR4963 & 13 09 56.99 & -05 32 20.43  &  4.38 & HR5793  \\
HD114961 &        & 13 14 04.45 & -02 48 24.70  &  7.02 & HR5867  \\
HD115604 & HR5017 & 13 17 32.54 & +40 34 21.38  &  4.72 & HR3314  \\
HD115659 & HR5020 & 13 18 55.29 & -23 10 17.44  &  3.00 & HR2421  \\
HD115892 & HR5028 & 13 20 35.81 & -36 42 44.26  &  2.70 & HR2421  \\
HD116656 & HR5054 & 13 23 55.54 & +54 55 31.30  &  2.70 & HR2421  \\
HD116658 & HR5056 & 13 25 11.57 & -11 09 40.75  &  1.04 & HR2421  \\
\hline
\end{tabular}
\end{table}

\begin{table}
\setcounter{table}{4}
%\caption{table4 contd..\\}
\begin{tabular}{cccccc} \hline
HD & HR & $\alpha$(J2000.0) &
$\delta$(J2000.0) &
V$_{mag}$ & Standard Star \\
& & & & & \\
(1) & (2) & (3) & (4) & (5) & (6) \\ \hline
HD116870 & HR5064 & 13 26 43.16 & -12 42 27.59  &  5.27 & HR5793  \\
HD120052 & HR5181 & 13 47 25.39 & -17 51 35.42  &  5.44 & HR5793  \\
HD120315 & HR5191 & 13 47 32.43 & +49 18 47.75  &  1.86 & HR2421  \\
HD121299 & HR5232 & 13 54 42.14 & -01 30 11.24  &  5.16 & HR5867  \\
HD123123 & HR5287 & 14 06 22.29 & -26 40 56.50  &  3.26 & HR3314  \\
HD123139 & HR5288 & 14 06 40.94 & -36 22 11.83  &  2.06 & HR5893  \\
HD123299 & HR5291 & 14 04 23.34 & +64 22 33.06  &  3.65 & HR2421  \\
HD123657 & HR5299 & 14 07 55.65 & +43 51 17.30  &  5.27 & HR5511  \\
HD123934 & HR5301 & 14 10 50.50 & -16 18 07.00  &  4.90 & HR4259  \\
HD124294 & HR5315 & 14 12 53.74 & -10 16 25.32  &  4.19 & HR2421  \\
HD126661 & HR5405 & 14 26 27.36 & +19 13 36.83  &  5.39 & HR3314  \\
HD127665 & HR5429 & 14 31 50.13 & +30 22 11.00  &  3.58 & HR6324  \\
HD129116 & HR5471 & 14 41 57.59 & -37 47 36.59  &  3.98 & HR5793  \\
HD129502 & HR5487 & 14 43 03.62 & -05 39 29.54  &  3.90 & HR5793  \\
HD130841 & HR5531 & 14 50 52.71 & -16 02 30.40  &  2.75 & HR5793  \\
HD130952 & HR5535 & 14 51 01.07 & -02 17 56.94  &  4.93 & HR5867  \\
HD131156 & HR5544 & 14 51 23.30 & +19 06 04.00  &  4.50 & HR6324  \\
HD131918 & HR5564 & 14 56 46.11 & -11 24 34.92  &  5.47 & HR5867  \\
HD134083 & HR5634 & 15 07 17.34 & +24 52 17.00  &  4.93 & HR5867  \\
HD135722 & HR5681 & 15 15 29.77 & +33 18 58.70  &  3.47 & HR6324  \\
HD136512 & HR5709 & 15 20 08.94 & +29 37 00.00  &  5.51 & HR5511  \\
HD138716 & HR5777 & 15 34 10.70 & -10 03 52.30  &  4.61 & HR5867  \\
HD138905 & HR5787 & 15 35 31.57 & -14 47 22.33  &  3.92 & HR2421  \\
HD141004 & HR5868 & 15 46 26.61 & +07 21 11.06  &  4.43 & HR6378  \\
HD141714 & HR5889 & 15 49 35.88 & +26 04 09.00  &  4.63 & HR5511  \\
HD141850 & HR5894 & 15 50 41.70 & +15 08 01.00  &  7.10 & HR6324  \\
HD145328 & HR6018 & 16 08 58.45 & +36 29 10.30  &  4.76 & HR5107  \\
HD147165 & HR6084 & 16 21 11.31 & -25 35 34.06  &  2.91 & HR6324  \\
HD147394 & HR6092 & 16 19 44.43 & +46 18 48.11  &  3.89 & HR5867  \\
HD148513 & HR6136 & 16 28 33.98 & +00 39 54.00  &  5.90 & HR6378  \\
HD149757 & HR6175 & 16 37 09.53 & -10 34 01.52  &  2.57 & HR5867  \\
\hline
\end{tabular}
\end{table}

\newpage
\begin{table}
%\caption{Physical Parameters of program stars}
\begin{tabular}{ccccccl} \hline
HD & Spectral & Luminosity &
T$_{eff}$($^{\circ}$K)& log$_{10}$(g) & (Fe/H) &Reference\\
 & Type & Class &
& & & \\
(1) & (2)& (3) & (4) & (5) & (6) & (7) \\ \hline
HD007927 & F0   & Ia    &       &      &       &         \\
HD008538 & A5   & III   & 8090  &      &       & 1995A\&AS...110..553 (Sokolov)  \\
HD010307 & G1.5 & V     & 5898  & 4.31 & -0.02 & 1993A\&A....275..101 (Edvardsson) \\
HD011353 & K0   & III   & 4600  & 2.70 & -0.13 & 1990ApJS....1075..1128 (McWilliam) \\
HD023475 & M2.5 & II    &       &      &       &         \\
HD025204 & B3   & V     &       &      &       &         \\
HD026846 & K3   & III   & 4582  & 2.70 & 0.21  & 1997A\&AS...124..299C (Cayrel) \\
HD030652 & F6   & V     & 6380  & 4.40 & 0.02  & 2004ApJS...152..251 (INDO-US) \\
HD030836 & B2   & III   & 22120 & 3.59 & -0.31 & 1997A\&AS...124..299C (Cayrel) \\
HD035468 & B2   & III   & 22570 & 3.72 & -0.25 & 2004ApJS...152..251 (INDO-US) \\
HD035497 & B7   & III   & 13622 & 3.80 & -0.10 & 2004ApJS...152..251 (INDO-US) \\
HD036673 & F0   & Ib    & 7400  & 1.10 & 0.04  & 2004ApJS...152..251 (INDO-US) \\
HD037128 & B0   & Iab   &       &      &       &         \\
HD037742 & O9   & Iab   &       &      &       &         \\
HD038393 & F7   & V     & 6398  & 4.29 & -0.07 & 1997A\&AS...124..299C (Cayrel) \\
HD038858 & G4   & V     &       &      &       &         \\
HD040136 & F1   & V     & 6939  & 4.23 & -0.13 & 2004ApJS...152..251 (INDO-US) \\
HD043232 & K1.5 & III   & 4270  & 2.22 & -0.18 & 2004ApJS...152..251 (INDO-US) \\
HD047105 & A0   & IV    & 9260  & 3.60 & -0.12 & 1994PASP...1239..1247 (Adelman) \\
HD047839 & O7   & Ve    &       &      &       &         \\
HD048329 & G8   & Ib    & 4150  & 0.80 & 0.20  & 2004ApJS...152..251 (INDO-US) \\
HD049331 & M1   & Iab   & 3600  & 0.70 & 0.17  & 1997A\&AS...124..299C (Cayrel) \\
HD054605 & F8   & Iab   & 6222  & 0.60 & 0.19  & 1981ApJ...1018..1034 (Luck)\\
HD054810 & K0   & III   & 4697  & 2.35 & -0.25 & 2004ApJS...151..387 (Ivanov) \\
HD056537 & A3   & V     &       &      &       &     \\
HD058715 & B8   & Ve    & 11710 &      &       & 1995A\&AS...110..553 (Sokolov)  \\
HD060414 & A4   & Ia    &       &      &       &         \\
HD061421 & F5   & IV    & 6650  & 4.10 & 0.04  & 1995PASP...219..224 (Andrievsky) \\
HD061935 & G9   & III   & 4776  & 2.20 & -0.03 & 2004ApJS...151..387 (Ivanov) \\
HD062345 & G8   & IIIa  & 5000  & 2.90 & -0.16 & 1990pJS....1075..1128 (McWilliam) \\
HD062576 & K3   & III   & 4308  & 1.30 & 0.01  & 1997A\&AS...124..299C (Cayrel) \\
HD062721 & K4   & III   & 3940  & 1.67 & -0.27 & 1997A\&AS...124..299C (Cayrel) \\
HD063700 & G6   & Ia    & 4990  & 1.15 & 0.24  & 1997A\&AS...124..299C (Cayrel) \\
HD066811 & O5   & Ia    & & & &  \\
HD067228 & G1   & IV    & 5779  & 4.20 & 0.04  & 2004ApJS...152..251 (INDO-US) \\
HD068312 & G6   & III   &       &      &       &     \\
HD070272 & K4.5 & III   & 3900  & 1.59 & -0.03 & 1997A\&AS...124..299C (Cayrel) \\
HD071369 & G5   & III   & 5300  & 2.67 & 0.06  & 2004ApJS...152..251 (INDO-US) \\
HD072094 & K5   & III   &       &      &       &     \\
HD074918 & G8   & III   & 4950  & 2.26 & -0.09 & 1997A\&AS...124..299C (Cayrel) \\
HD076943 & F3   & V     & 6590  & 4.00 & 0.25  & 2004ApJS...152..251 (INDO-US) \\
HD077912 & G7   & Ib-II & 5000  & 2.00 & 0.38  & 2004ApJS...152..251 (INDO-US) \\
HD080874 & M0   & III   & & & &  \\
HD081797 & K3   & II    & 4120  & 1.77 & -0.12 & 1990ApJS....1075..1128 (McWilliam)  \\
HD082328 & F6   & IV    & 6380  & 4.09 & -0.20 & 1993A\&A...101..152 (Edvardsson) \\
HD084748 & M8   & IIIe  & & & &  \\
HD085444 & G6   & III   & 5000  & 2.93 & -0.14 & 2004ApJS...152..251 (INDO-US) \\
\hline
\end{tabular}
\end{table}

\begin{table}
\setcounter{table}{5}
%\caption{table5 contd..\\}
\begin{tabular}{ccccccl} \hline
HD & Spectral & Luminosity &
T$_{eff}$($^{\circ}$K)& log$_{10}$(g) & (Fe/H) &Reference\\
 & Type & Class &
& & & \\
(1) & (2)& (3) & (4) & (5) & (6) & (7) \\ \hline

HD085951 & k5   & III   &       &      &       &        \\
HD086663 & M2   & III   &       &      &       &        \\
HD087737 & A0   & Ib    & 9700  & 2.00 & -0.05 & 1995ApJS...659..692 (Venn)  \\
HD088230 & K8   & V     & 4000  & 4.50 & 0.28  & 1997A\&AS...124..299C (Cayrel)\\
HD088284 & K0   & III   & 4971  & 2.70 & 0.39  & 2004ApJS...151..387 (Ivanov)\\
HD089025 & F0   & III   &       &      &       &    \\
HD089021 & A2   & IV    & 9280  & 3.90 & 0.20  & 1995A\&A...536..546 (Hill)\\
HD089449 & F6   & IV    & 6333  & 4.06 & 0.21  & 2004ApJS...152..251 (INDO-US)\\
HD089490 & K0   &       &       &      &       &     \\
HD089758 & M0   & III   &       &      &       &      \\
HD090254 & M3   & III   & 3706  & 1.40 & 0.11  & 1997A\&AS...124..299C (Cayrel) \\
HD090432 & K4   & III   & 3950  & 1.68 & -0.12 & 1997A\&AS...124..299C (Cayrel)\\
HD090610 & K4   & III   & 3990  & 1.77 & -0.39 & 1997A\&AS...124..299C (Cayrel) \\
HD092125 & G2.5 & IIa   & 5600  & 2.10 & 0.38  & 2004ApJS...152..251 (INDO-US)\\
HD092588 & K1   & IV    & 5044  & 3.60 & -0.10 & 2004ApJS...152..251 (INDO-US)\\
HD093813 & K0   & III   & 4250  & 2.32 & -0.24 & 2004ApJS...152..251 (INDO-US) \\
HD094264 & K0   & III   & 4670  & 2.96 & -0.20 & 2004ApJS...152..251 (INDO-US)\\
HD094481 & K0   & II+.. &       &      &       &    \\
HD095418 & A1   & V     & 9620  & 3.90 & 0.16  & 2004ApJS...152..251 (INDO-US)\\
HD097603 & A4   & V     & 8080  &      &       & 1995A\&AS...110..553 (Sokolov) \\
HD097778 & M3   & IIb   & 3300  &      & 0.00  & 2004ApJS...151..387 (Ivanov)\\
HD098231 & G0   & V     & 5950  & 4.30 & -0.35 & 1994A\&A...505..516 (Cayrel) \\
HD099028 & F1   & IV    & 6739  & 3.98 & 0.06  & 2004ApJS...152..251 (INDO-US)\\
HD099167 & K5   & III   & 3930  & 1.61 & -0.38 & 2004ApJS...152..251 (INDO-US)\\
HD100920 & G8.5 & III   & 4800  & 2.93 & -0.34 & 2004ApJS...152..251 (INDO-US) \\
HD101501 & G8   & V     & 5360  & 4.35 & -0.39 & 2004ApJS...151..387 (Ivanov)\\
HD102212 & M1   & III   & & & &  \\
HD105707 & K2   & III   & 4320  & 2.16 & -0.13 & 1997A\&AS...124..299C (Cayrel) \\
HD106625 & B8   & III   &       &      &       &        \\
HD107259 & A2   & IV    & 9333  & 3.00 & 0.11  & 2004ApJS...152..251 (INDO-US)\\
HD107328 & K0   & IIIb  & 4380  & 2.39 & -0.48 & 2004ApJS...152..251 (INDO-US)\\
HD109358 & G0   & V     & 5903  & 4.42 & -0.12 & 2004ApJS...151..387 (Ivanov)\\
HD109379 & G5   & II    & 5170  & 2.10 & -0.11 & 1997A\&AS...124..299C (Cayrel)\\
HD110379 & F0   &  V    & 7099  & 4.00 & -0.57 & 1997A\&AS...124..299C (Cayrel)\\
HD111812 & G0   & IIIp  &       &      & 0.01  & 2004ApJS...152..251 (INDO-US)\\
HD112142 & M3   & III   & & & &  \\
HD112300 & M3   & III   & 3652  & 1.3  & -0.09 & 1985ApJ...326..338 (Smith) \\
HD113139 & F2   & V     &       &      &       &   \\
HD113226 & G8   & III   & 4994  & 2.10 & 0.12  & 2004ApJS...151..387 (Ivanov)\\
HD113847 & K1   & III   & 4510  & 2.20 & -0.09 & 2004ApJS...152..251 (INDO-US)\\
HD113996 & K5   & III   & 3970  & 1.69 & -0.26 & 2004ApJS...151..387 (Ivanov)\\
HD114330 & A1   & Vs+.. & 9509  & 3.60 & -0.02 & 2004ApJS...152..251 (INDO-US)\\
HD114961 & M7   & III   & 3014  & 0.40 & -0.81 & 2004ApJS...152..251 (INDO-US)\\
HD115604 & F3   & III   & 7200  & 3.00 &  0.18 & 2004ApJS...152..251 (INDO-US)\\
HD115659 & G8   & III   & 5025  & 2.60 &  0.06 & 1995AJ...2968..3009 (Luck)\\
HD115892 & A2   & V     & 9030  &      &       & 1995A\&AS...110..553 (Sokolov) \\
HD116656 & A2   & V     & 5793  &      &       & 2004ApJS...152..251 (INDO-US)\\
HD116658 & B1   & III   &       &      &       &        \\
HD116870 & K5   & III   &       &      &       &        \\
\hline
\end{tabular}
\end{table}

\begin{table}
\setcounter{table}{5}
%\caption{table5 contd..\\}
\begin{tabular}{ccccccl} \hline
HD & Spectral & Luminosity &
T$_{eff}$($^{\circ}$K)& log$_{10}$(g) & (Fe/H) &Reference\\
 & Type & Class &
& & & \\
(1) & (2)& (3) & (4) & (5) & (6) & (7) \\ \hline

HD120052 & M2 & III & & & &  \\
HD120315 & B3 & V   & 17200 &      &       & 1995A\&AS...110..553 (Sokolov) \\
HD121299 & K2 & III & 4710  & 2.64  & 0.03 & 1990ApJS...1075..1128 (McWilliam)  \\
HD123123 & K2 & III & 4600  & 2.00  & -0.06& 1991ApJS....579..  (Luck) \\
HD123139 & K0   & IIIb  & 4980  & 2.75 & 0.03  & 1997A\&AS...124..299C (Cayrel)\\
HD123299 & A0   & III   & 10080 & 3.30 & -0.56 & 2004ApJS...152..251 (INDO-US)\\
HD123657 & M4.5 & III   & 3452  & 0.90 & -0.03 & 2004ApJS...152..251 (INDO-US)\\
HD123934 & M2   & IIIa  &       &      &       &        \\
HD124294 & K2.5 & IIIb  & 4120  & 2.06 & -0.39 & 1997A\&AS...124..299C (Cayrel)\\
HD126661 & F0m  &       & 7754  & 3.50 & 0.10  & 2004ApJS...152..251 (INDO-US)   \\
HD127665 & K3   & III   & 4260  & 2.22 & -0.17 & 2004ApJS...152..251 (INDO-US) \\
HD129116 & B3   & V     &       &      &       &        \\
HD129502 & F2   & III   & 6820  &      &       & 1995A\&AS...110..553 (Sokolov) \\
HD130841 & A3   & IV    &       &      &       &        \\
HD130952 & G8   & III   & 4820  & 2.91 & -0.39 & 1990ApJS...1075..1128 (McWilliam) \\
HD131156 & G7   & Ve    & 5500  & 4.60 & -0.15 & 2004ApJS...152..251 (INDO-US) \\
HD131918 & K4   & III   & 3970  & 1.72 & 0.22  & 1990ApJS...1075..1128 (McWilliam) \\
HD134083 & F5   & V     & 6632  & 4.50 & 0.32  & 2004ApJS...152..251 (INDO-US) \\
HD135722 & G8   & III   & 4834  & 2.45 & -0.39 & 2004ApJS...151..387 (Ivanov)\\
HD136512 & K0   & III   & 4730  & 2.75 & -0.44 & 2004ApJS...152..251 (INDO-US) \\
HD138716 & K1   & IV    & 4730  & 3.20 & -0.12 & 1990ApJS...1075..1128 (McWilliam) \\
HD138905 & K0   & III   & 4700  & 3.01 & -0.42 & 1990ApJS...1075..1128 (McWilliam)  \\
HD141004 & G6   & V     & 5937  & 4.21 & -0.04 & 2004ApJS...152..251 (INDO-US) \\
HD141714 & G5   & III   & 5230  & 3.15 & -0.32 & 2004ApJS...152..251 (INDO-US) \\
HD141850 & M7   & III   &       &      &       &        \\
HD145328 & K1   & III   & 4678  & 2.50 & -0.14 & 2004ApJS...151..387 (Ivanov)\\
HD147165 & B1   & III   &       &      &       &        \\
HD147394 & B5   & IV    & 15000 & 3.95 & 0.15  & 1993A\&A...335..355 (Smith)\\
HD148513 & K4   & III   & 4046  & 1.00 & 0.25  & 2004ApJS...151..387 (Ivanov) \\
HD149757 & O9   & V     & & & &  \\ \hline
\end{tabular}
\end{table}

In conclusion, we may mention that this library of 126 stellar
spectra in the NIR J band has been carefully checked for its
consistency with earlier published libraries and provides a larger
database with extended spectro-luminosity coverage for usage in
stellar population synthesis work and other applications as well as
complimenting large optical libraries.

\section*{Acknowledgments}
The research was partly funded by a grant from ISRO RESPOND to HPS.
The research work at the Physical Research Laboratory is funded by
the Department of Space, Government of India. This paper has made
use of the SIMBAD database, operated at CDS, Strasbourg, France.

\end{document}